\documentstyle[aps,eqsecnum,epsfig]{revtex}

\widetext

\begin{document}
\title{Phase noise measurement in a cavity with a movable mirror
undergoing quantum Brownian motion}
\author{Vittorio Giovannetti and David Vitali}
\address {Dip. di Matematica e Fisica and Unit\`a INFM, Universit\`a di
Camerino, via Madonna delle Carceri 62032, Camerino, Italy}
\date{\today}
\maketitle

\begin{abstract}
We study the dynamics of an optical mode in a cavity with a movable 
mirror subject to quantum Brownian motion.
We study the phase noise power spectrum of the 
output light, and we describe the mirror Brownian motion, which is 
responsible for the thermal noise contribution, using the 
quantum Langevin approach. We show that the standard quantum Langevin 
equations, supplemented with the appropriate non-Markovian
correlation functions, provide an adequate description of Brownian
motion.  
\end{abstract}


\section{Introduction}

The mechanical interaction between a moving mirror and a radiation field
has been an important topic for the study of very high precision 
optical interferometers in which radiation pressure effects cannot be ignored.
This interaction is at the basis of the interferometric detection
of gravitational waves, where the tiny displacement of a mirror
can be detected as a phase shift of the interference fringes
\cite{grav}. Another interesting application is the atomic force microscope
\cite{afm}, where an image of a surface at atomic resolution is obtained from the 
measurement of the force between the surface and a probe tip mounted on
a microcantilever. 

A cavity with a movable mirror is of interest also for cavity QED studies, 
which usually involves the quantum coherent interaction between high-Q
cavity modes at low photon number and single atoms. In this case,
the atomic degrees of freedom are replaced by the motional degree
of freedom of the movable mirror. Interesting quantum effects, 
as the generation of
sub-Poissonian light \cite{subp}, of Schr\"odinger cat states of both
the cavity mode \cite{cat1} and even of the mirror \cite{cat2}
have been already illustrated.

In these applications one needs a very high resolution for position 
measurements and a good control of the various noise sources, because one has
to detect the effect of a very weak force. As shown by the pioneering work
of Braginsky \cite{brag}, even though all classical noise sources
had been minimized, the detection of gravitational waves would be ultimately
determined by quantum fluctuations and the Heisenberg uncertainty principle.
Quantum noise in interferometers has two fundamental sources, 
the photon shot noise 
of the laser beam, prevailing at low laser intensity, and the fluctuations
of the mirror position due to radiation pressure, which is 
proportional to the incident laser power. This radiation pressure noise is the
so-called ``back-action noise'' arising from the fact that intensity fluctuations
affect the momentum fluctuations of the mirror, which are then fed back into
the position by the dynamics of the mirror. The two quantum noises are
minimized at an optimal, intermediate, laser power, yielding the so-called
{\em standard quantum limit} (SQL),
which coincides with mean square fluctuations
of the harmonic oscillator ground state $\Delta q_{SQL} = \sqrt{\hbar/2m
\omega_{\cal S}}$ ($\omega_{\cal S}$ is the mirror oscillation frequency).
Real devices constructed up to now are still far from the standard
quantum limit because quantum noise is much smaller than that of classical
origin, which is essentially given by thermal noise. In fact, present
interferometric gravitational wave detectors are limited by the Brownian
motion of the suspended mirrors \cite{abra}, which can be decomposed into 
suspension and internal (i.e. of internal acoustic modes) thermal noise.
Therefore it is very important to establish the experimental limitations
determined by thermal noise and recent experiments \cite{hadjar,titto}
have obtained interesting results. With this respect it is also 
important to establish which is the most appropriate formal description of 
quantum Brownian motion. In fact, even though the classical 
understanding of the phenomenon is well established, relying on 
Langevin or Fokker-Planck equations \cite{risken}, 
its quantum generalization is 
still the subject of an intense debate (see \cite{vacchi} and 
references therein). In particular, the recent paper by 
Jacobs {\it et al.} \cite{JACOB} has shown that the standard description of 
quantum Brownian motion, which is the straightforward generalization
of the classical case \cite{caldleg},
gives an inadequate description
since it generates a non-sensical term in the optical phase noise spectrum
in the case of a cavity with a movable mirror.
The authors of \cite{JACOB} adopt therefore a corrected 
quantum Langevin equation,
based on the Di\`osi master equation \cite{diosi1}, and suggest that, 
even if it is quite challenging, 
the corresponding modifications of the phase noise spectrum could be
revealed using miniature high-frequency mechanical oscillators and 
ultra-low temperatures.
In the present paper we shall reconsider the same system, i.e. a driven 
cavity with a movable mirror, and shall show that the inadequacy shown 
in \cite{JACOB} has to be traced back to the inadequacy of the 
quantum noise commutation relations and correlation functions
which are dictated by the standard Brownian motion master equation. 
We shall see that, differently from the master equation 
approach, a consistently applied quantum Langevin equation 
\cite{qnoise} provides a flexible approach, valid at {\em any
temperature} and therefore also in the fully quantum regime of very low
temperatures. This however does not mean that the quantum Langevin equation
approach is generally superior than the master equation approach, 
but simply that
in the case under study, which is a linearized, non-markovian problem,
the quantum Langevin description is more convenient and powerful. 

The outline of the paper is as follows. In Sec. II the 
appropriate quantum Langevin 
equations for a Brownian particle are derived starting from the usual model 
based on the coupling with a reservoir of harmonic oscillator, and its 
consistency is shown. In Sec. III the quantum Langevin approach 
is applied to the case of a cavity mode with a movable mirror
and the homodyne spectrum of the reflected light,
showing the thermal and 
quantum fluctuations of the mirror, is studied. Sec. IV is for 
concluding remarks.

\section{The dynamics of the system}

The system studied in the present paper consists of a coherently 
driven optical cavity with a moving mirror. This opto-mechanical 
system can represent one arm of an interferometer able to detect weak 
forces as those associated with gravitational waves \cite{grav} or 
an atomic force microscope \cite{afm}. 
The detection 
of very weak forces requires having quantum limited devices, whose 
sensitivity is ultimately determined by the quantum fluctuations.
For this reason we shall describe the mirror as a {\em quantum} 
mechanical harmonic oscillator with mass $m$ and frequency 
$\omega_{\cal S}$. 
The optomechanical coupling between the mirror and
the cavity field is realized by the radiation pressure. The 
electromagnetic field exerts a force on the movable mirror which is 
proportional to the intensity of the field, which, at the same time, 
is phase-shifted by $2kq$, where $k$ is the wave vector and $q$ is
the mirror displacement from the equilibrium position.
In the adiabatic limit in which the mirror frequency is much smaller 
than the cavity free spectral range $c/2L$ ($L$ is the cavity length)
\cite{law}, one can focus on one cavity mode only because
photon scattering into other modes can be neglected, and one has the 
following Hamiltonian
\begin{equation}
	H=\hbar \omega_{c}b^{\dagger}b + \frac{p^{2}}{2m}+
	\frac{1}{2} m \omega_{\cal S}^{2}q^{2} -\hbar \frac{\omega_{c}}{L} 
	q b^{\dagger}b + i\hbar E\left(b^{\dagger}e^{-i\omega_{0}t}-b
	e^{i\omega_{0}t}\right) \;, 
	\label{hparte}
\end{equation}
where $b$ is the cavity mode annihilation operator with optical
frequency $\omega_{c}$ and $E$ describes the coherent input field
with frequency $\omega_{0}\sim \omega_{c}$
driving the cavity. The quantity $E$ is related to the input laser 
power $P$ by $E=\sqrt{P\gamma_{c}/\hbar \omega_{0}}$, where 
$\gamma_{c}$
is the cavity decay constant due to the input coupling mirror.
Since we shall focus on the quantum and thermal noise of the system,
we shall neglect all the experimental sources of noise, i.e., 
we shall assume that the driving laser is stabilized
in intensity and frequency. This means neglecting all
the fluctuations of the complex parameter $E$. Including these 
supplementary noise sources is however quite straightforward and a 
detailed calculation of their effect is shown in Ref.~\cite{JACOB}. 
Moreover recent experiments have shown
that classical laser noise can be made negligible in the relevant 
frequency range \cite{hadjar,titto}.
The adiabatic regime $\omega_{\cal S} \ll c/2L$ we have assumed 
in Eq.~(\ref{hparte}) implies $\omega_{\cal S} \ll \omega_{c}$,
and therefore the generation of photons due to the Casimir effect,
and also retardation and Doppler effects are completely negligible.

The dynamics of the system is not only determined by the Hamiltonian
interaction (\ref{hparte}), but also by the dissipative interaction
with external degrees of freedom. The cavity mode is damped due to the
photon leakage through the mirrors which couple the cavity 
mode with the continuum of the outside electromagnetic modes. For simplicity 
we assume that the movable mirror has perfect reflectivity and that
transmission takes place through the other ``fixed'' mirror only (see
Fig.~1 for a schematic description of the system). 
The mechanical oscillator, which may represent not only the
center-of-mass degree of freedom of the mirror, but also a torsional 
degree of freedom as in \cite{titto}, or an internal acoustic mode
as in \cite{hadjar}, undergoes Brownian motion caused by the 
uncontrolled coupling with other internal and external modes at the 
equilibrium temperature $T$.

The dissipative dynamics of the optical cavity mode is well described
by the so-called vacuum optical master equation \cite{milwal}
\begin{equation}
\label{vome}
\dot{\rho }= \frac{\gamma_{c}}{2}
\left(2b\rho b^{\dagger}-b^{\dagger}b\rho 
-\rho b^{\dagger}b\right) \;,
\end{equation}
for the time evolution of the density matrix of the whole system $\rho$.
In fact, the mean thermal number of photons at the optical frequency $\omega_c$
is extremely small and thermal excitation is therefore completely 
negligible. The time evolution generated by Eq.~(\ref{vome}) 
presents no ambiguity. In fact it is of Lindblad form \cite{lind} and 
therefore it preserves the positivity of the density matrix. Moreover
it is completely equivalent to the time evolution for the operators
in the Heisenberg representation
driven by the following quantum Langevin equation
\begin{equation}
\label{qle}
\dot{b}(t)= - \frac{\gamma_{c}}{2}b(t)+\sqrt{\gamma_{c}}b_{in}(t)\;,
\end{equation}
where $b_{in}(t)$ is the input noise operator associated with the vacuum 
fluctuations of the continuum of modes outside the cavity, having the 
following commutation relation
\begin{equation}
\left[b_{in}(t),b_{in}^{\dagger}(t')\right]=\delta(t-t')
\label{commu}
\end{equation}
and correlation functions
\begin{eqnarray}
&& \langle b_{in}(t)b_{in}(t') \rangle = \langle 
b_{in}^{\dagger}(t)b_{in}(t') \rangle
= 0 \label{correb1}\\
&& \langle b_{in}(t)b_{in}^{\dagger}(t') \rangle = \delta(t-t')  
\label{correb2}\;. 
\end{eqnarray}

The description of the quantum Brownian motion of a massive particle
in a potential is instead not so well established. The 
standard Brownian motion master equation (SBMME) has been first derived by
Caldeira and Leggett \cite{caldleg} in the high temperature limit
and reads
\begin{equation}
\dot{\rho}(t)=-\frac{i}{\hbar}\left[H_{\cal S},\rho\right]
-\frac{i\eta}{2 m \hbar}\left[q(t),\{ p(t), \rho(t) \}\right] 
-\frac{\eta kT}{\hbar^{2}}
\left[q(t),[ q(t), \rho(t) ] \right],
\label{cl}
\end{equation}
where $H_{\cal S}$ is the uncoupled particle Hamiltonian, $p$ is its 
momentum and $\eta \dot{q}$ is the friction force. This master equation
is the direct generalization of the Fokker-Planck equation for the 
classical Brownian motion but, since it is not of
Lindblad form, it has the drawback that it does not
ensure the positivity of the density operator \cite{ambe}.
This fact has stimulated many authors who have amended the SBMME
with additional terms so to cast it into the Lindblad form 
\cite{vacchi,diosi1,diosi2,gao}. These corrected master equations 
preserve the positivity of the density matrix but also them are valid
in the high temperature limit only, as the SBMME, 
because they necessarily provide a Markovian description of Brownian 
motion which instead becomes highly non-Markovian 
in the low temperature limit \cite{haake}. In fact, in this limit, 
the reservoir correlation time is no more negligible 
because it is essentially determined by the 
``thermal time'' $\tau_{T} = \hbar/kT$.

An alternative description of quantum Brownian motion is provided by 
the quantum Langevin equations for the Heisenberg operators $q(t)$ 
and $p(t)$, which, in 
analogy with the classical case, should read 
\begin{eqnarray}
\dot{q}(t)&=&\frac{i}{\hbar} \left[H_{\cal S},q(t)\right] 
\label{qleqbm0}\\
\dot{p}(t)&=&\frac{i}{\hbar}\left[H_{\cal 
S},p(t)\right]-\frac{\eta}{m}p(t) +\xi(t) \;,
\label{qleqbm}
\end{eqnarray}
where $\xi(t)$ is a Hermitian noise operator with correlation function
\begin{equation}
\langle \xi(t) \xi(t')\rangle =2\eta kT \delta(t-t') \;.
\label{correwro}
\end{equation}
One would expect that these equations give correct results at least 
in the high temperature limit, where the classical limit should be 
recovered. Instead Ref.~\cite{JACOB} has shown 
that they give inconsistent results even in this limit
because they do not preserve the 
commutation relation $\left[q(t), p(t)\right]=i\hbar$ and they yield 
a spurious term in the phase fluctuation spectrum. In fact, in spectral 
measurements, the Fourier transform of 
correlation functions of the form $G(\tau)=\langle 
R(t)R(t+\tau)\rangle $ are measured, where $R(t)$ is an appropriate 
output field. This correlation function depends only on $\tau$ because 
of stationarity, and moreover it is an even function of $\tau$ because
$R(t)$ commutes with itself at different times. This implies that the 
observed spectrum has to be an {\em even} function of the frequency
$\omega$, while the adoption of the quantum Langevin equations 
(\ref{qleqbm0})-(\ref{qleqbm}) yields a term which is an {\em odd} function of 
$\omega $ \cite{JACOB}.

In Ref.~\cite{JACOB} these inadequacies of the quantum Langevin
equations (\ref{qleqbm0})-(\ref{qleqbm}) have been traced back
to the fact that they are equivalent to the 
SBMME (\ref{cl}), which is not of Lindblad form and therefore
does not preserve positivity. For this reason
they consider the amended master equation of the Lindblad
form proposed by Di\'osi in
\cite{diosi1}, and derive the set of quantum Langevin equation equivalent to
it, 
\begin{eqnarray}
\label{qledio0}
\dot{q}(t)&=&\frac{i}{\hbar} \left[H_{\cal S},q(t)\right] +\epsilon (t)\\
\dot{p}(t)&=&\frac{i}{\hbar}\left[H_{\cal 
S},p(t)\right]-\frac{\eta}{m}p(t) +\xi(t) \;,
\label{qledio}
\end{eqnarray}
having the additional noise term $\epsilon (t)$.
The corresponding correlation functions are
\begin{eqnarray}
\langle \xi(t) \xi(t')\rangle &=& 2\eta kT \delta(t-t')
\label{correwro1} \\
\langle \epsilon(t) \epsilon(t')\rangle &=& \frac{\hbar^2 \eta}{6m^2 kT}
 \delta(t-t')
\label{correwro2} \\
\langle \xi(t) \epsilon(t')\rangle &=& -i\hbar \frac{\eta}{2m} \delta(t-t')
\label{correwro3} \\
\langle \epsilon(t) \xi(t')\rangle &=& i\hbar \frac{\eta}{2m} \delta(t-t')
\label{correwro4} \;.
\end{eqnarray}
It is then possible to see that the phase noise spectrum associated
with this dynamical description of quantum Brownian motion has no spurious term
and that it is an even function of the frequency, as it must be.
However, the approach of Ref.~\cite{JACOB} 
can be questioned for two reasons. First of all
the master equation of Ref.~\cite{diosi1} has been derived by Di\'osi
with heuristic arguments and in a later paper \cite{diosi2} Di\'osi
himself corrected it by considering a more rigorous medium
temperatures extension of the SBMME. The master equation
of Ref.~\cite{diosi2} would lead to a different
set of quantum Langevin equations; more generally speaking, the Lindblad 
form condition does not uniquely determine the master equation, and 
therefore the form of the quantum Langevin equations neither. Furthermore,
the added noise term $\epsilon (t)$ in Eq.~(\ref{qledio0})
is not present in the classical case and it has an unclear physical origin.

For this reason we reconsider here the problem, assuming
a different starting point. Most of the derivations of the Brownian
motion master equations are based on the independent oscillator model
for the reservoir, whose quite general validity has been extensively
discussed in \cite{caldleg2}. Therefore, rather than first deriving
the master equation from this reservoir model and then
considering the quantum Langevin equations associated to it, we 
derive the quantum Langevin equations directly from the reservoir oscillator
model, as it is shown in \cite{qnoise,ocon}.

Let us neglect for the moment the presence of the cavity mode and 
consider a particle undergoing Brownian motion, with Hamiltonian
$H_{\cal S}$. The reservoir is described by a collection of independent
harmonic oscillators with frequency $\omega_j$, couplings
$k_j$, and whose canonical coordinates $q_j$ and $p_j$ have
been appropriately rescaled. The total system Hamiltonian is 
\cite{caldleg,qnoise,ocon}
\begin{equation}
H_{\cal S} + \frac{1}{2}\sum_{j}( (p_{j}-k_{j}q)^{2} + \omega_{j}^{2} 
q_{j}^{2} ).
\label{primo1}
\end{equation}
The quantum Langevin equations can be obtained from the Heisenberg 
equations for $q(t)$, $p(t)$ and the reservoir annihilation 
operators $a_j(t)= \left(\omega_j q_j(t)+ip_j(t)\right)/\sqrt{2\hbar \omega_j}$,
\begin{eqnarray}
  \dot{q}(t) &=& \frac{i}{\hbar} \left[H_{\cal S},q(t)\right]\label{secondo1} \\
  \dot{p}(t) &=& \frac{i}{\hbar} \left[H_{\cal S},p(t)\right] 
  +\sum_{j}k_{j}\left(p_{j}(t)-k_{j}q(t)\right)
  \label{secondo2}\\
  \dot{a_{j}}(t) &=& -i \omega_{j}a_{j}(t) - k_{j} 
  \sqrt{\frac{\omega_{j}}{2\hbar}} q(t) \; . \label{secondo3}
\end{eqnarray} 
If we integrate the equation for 
$a_{j}(t)$ starting from the initial time $t_{0}$, we get 
\begin{equation}
a_{j}(t)=e^{-i \omega_{j}(t-t_{0})} a_{j}(t_{0}) - 
k_{j}\sqrt{\frac{\omega_{j}}{2 \hbar}} \int_{t_{0}}^{t} dt^{\prime} 
e^{-i \omega_{j}(t-t^{\prime})}q(t^{\prime}),
\label{secondo4}
\end{equation}
which, using integration by parts and 
the fact that $\dot{q}=\frac{i}{\hbar} \left[H_{\cal S},q\right]=p/m$, 
can be rewritten as 
\begin{equation}
a_{j}(t)=ik_{j}\sqrt{\frac{1}{2\hbar \omega_{j}}}q(t)+
e^{-i \omega_{j}(t-t_{0})} a_{j}(t_{0}) - 
ik_{j}\sqrt{\frac{1}{2 \hbar \omega_{j}}} e^{-i \omega_{j}(t-t_{0})} q(t_{0})
-i\frac{k_{j}}{m}\sqrt{\frac{1}{2 \hbar \omega_{j}}} \int_{t_{0}}^{t} dt^{\prime} 
e^{-i \omega_{j}(t-t^{\prime})}p(t^{\prime}).
\label{secondo5bis}
\end{equation}
Then we replace (\ref{secondo5bis}) in the equation for $\dot{p}(t)$, 
(\ref{secondo2}), which becomes
\begin{equation}
\dot{p}(t) = \frac{i}{\hbar} \left[H_{\cal S},p(t)\right] 
  + \tilde{Q}(t)-\sum_{j}k_{j}^{2}\cos\left[\omega_{j}(t-t_{0})\right]q(t_{0})
  - \int_{t_{0}}^{t} dt^{\prime}
\sum_{j}k_{j}^{2}\cos\left[\omega_{j}(t-t^{\prime})\right]
\frac{p(t^{\prime})}{m} ,
  \label{secondo6bis}
\end{equation}
where we have defined the reservoir operator
\begin{equation}
\tilde{Q}(t)=i\sum_{j}k_{j}\sqrt{\frac{\hbar \omega_{j}}{2}}\left(
e^{i \omega_{j}(t-t_{0})} a_{j}^{\dagger}(t_{0})
-e^{-i \omega_{j}(t-t_{0})} a_{j}(t_{0})\right).
\label{qu}
\end{equation}
As it is well known, the irreversible properties of the reservoir are
obtained only when an infinite number of oscillators, distributed
over a continuum of frequencies, is considered. The continuous limit
has to be performed according to the following prescription
\begin{equation}
\sum_{j} k_{j}^{2} \cdots \rightarrow \int_{0}^{\Omega} d \omega \, 
k^{2}(\omega) \, \frac{ d n }{ d \omega} \cdots \,= \,\frac{2 
\eta}{\pi}\int_{0}^{\Omega} d \omega \cdots,
\label{primo15}
\end{equation}
where $dn/d\omega $ is the oscillators density, $\eta$ is 
just the friction coefficient, and $\Omega $ is the frequency cutoff 
of the reservoir oscillator spectrum. When the continuous limit is
considered, the Heisenberg equations for the Brownian particle
become
\begin{eqnarray}
  \dot{q}(t) &=& \frac{p(t)}{m} \label{sblib}\\
  \dot{p}(t) &=& \frac{i}{\hbar} \left[H_{\cal S},p(t)\right]+
  \tilde{Q}(t) - 
  2 \eta \tilde{\delta}(t-t_{0}) q(t_{0}) - 
  2 \eta  \int_{t_{0}}^{t} d t^{\prime} 
  \tilde{\delta}(t-t^{\prime}) \frac{p(t^{\prime})}{m}  \; ,
\label{secondo5}
\end{eqnarray}
where we have defined 
the following function
\begin{equation}
\tilde{\delta}(t)=\frac{1}{\pi}\int_{0}^{\Omega} d \omega 
\cos(\omega t).
\label{primo19}
\end{equation}
Eqs.~(\ref{sblib}) and (\ref{secondo5}) become identical to the usual Langevin 
equations (\ref{qleqbm0})-(\ref{qleqbm}) 
when the usual assumption of a reservoir dynamics much faster
than that of the Brownian particle is made. This means
making a coarse-grained description in time equivalent to assuming 
the infinite cutoff limit $\Omega \rightarrow \infty $, under
which the function $\tilde{\delta}(t)$ becomes a Dirac delta
function. The reservoir operator $\tilde{Q}(t)$ in 
Eq.~(\ref{secondo5}) plays therefore the 
role of the random Langevin force, commuting 
with a generic
system operator evaluated at the initial time 
$t_{0}$, for every value of $t$. This fact suggests to interpret 
$\tilde{Q}(t)$ as the input noise of the system and the above 
quantum Langevin equations in terms of the 
input-output formalism developed by Gardiner and Collett \cite{io}. 
However the interpretation of
$\tilde{Q}(t)$ as an input noise must be made with care, because
{\em its commutation relations are different from those
of the typical input noise 
operators} (see Eq.~(\ref{commu})). In fact, using definition
(\ref{qu}) and the continuous
limit  prescription (\ref{primo15}), one derives the following 
commutation relation
\begin{equation}
\left[ \tilde{Q}(t),\tilde{Q}(t^\prime) \right]
=2i \hbar \eta \frac{d}{dt} \tilde{\delta}(t-t^{\prime}), 
\label{secondo6}
\end{equation}
which is not a delta function in $(t-t^{\prime})$, even in
the coarse-grained time limit $\Omega \rightarrow \infty$.
With this respect, it is interesting to consider also the correlation 
function of the noise operator $\tilde{Q}(t)$, which is 
generally defined as a trace over the reservoir degres of 
freedom,
\begin{equation}
\langle \tilde{Q}(t)\tilde{Q}(t^\prime) \rangle_{\cal B}
=tr_{\cal B} \{\tilde{Q}(t)\tilde{Q}(t^\prime) R_{0} \}, 
\label{primo14}
\end{equation}
where $R_{0}$ is the density operator of the 
reservoir at thermal equilibrium at temperature $T$,
\begin{equation}
R_{0}=\prod_{j}e^{-\frac{\hbar \omega_{j} a_{j}^{\dagger} a_{j}}{k 
T}}(1-e^{-\frac{\hbar \omega_{j}}{k T}}).
\label{primo12}
\end{equation}
Using again Eqs.~(\ref{qu}) and (\ref{primo15}), one gets
\begin{equation}
\langle \tilde{Q}(t)\tilde{Q}(t^\prime) \rangle_{\cal B}=
\frac{\hbar \eta}{\pi} \Big\{ {\cal F}_{r}(t-t^\prime) + i 
{\cal F}_{i}(t-t^\prime) \Big\},
\label{primo16}
\end{equation}
with
\begin{eqnarray}
{\cal F}_{r}(\tau)&=&\int_{0}^{\Omega} d\omega \;
  \omega \cos(\omega \tau) \coth\left(\frac{\hbar\omega}{2 k T} \right) 
  \label{primo17} \\
{\cal F}_{i}(\tau)&=& - \int_{0}^{\Omega} d\omega \;
  \omega \sin(\omega \tau) = \pi \frac{d}{d\tau}\tilde{\delta}(\tau),
  \label{primo18}  
\end{eqnarray}
The antisymmetric part, corresponding to ${\cal F}_{i}$, is a direct
consequence of the commutation relations (\ref{secondo6}) and, as we 
have seen, is never a Dirac delta, while the symmetric part, 
corresponding to ${\cal F}_{r}$, explicitely depends on temperature 
and becomes proportional to a Dirac delta function only if the 
high temperature limit $ kT \gg \hbar \Omega $ first, and the infinite frequency 
cutoff limit $\Omega \rightarrow \infty$ later, are taken. 
Eqs.~(\ref{secondo6}) and (\ref{primo16})-(\ref{primo18})
show the non-Markovian nature of quantum Brownian motion, which 
becomes particularly evident in the low temperature limit \cite{haake}.
Therefore, 
assuming the independent oscillator model for the reservoir, which is the usual 
starting point for the derivation of the Brownian motion master 
equation, we have derived the {\em exact} quantum Langevin equations
(\ref{sblib}) and (\ref{secondo5}), which reduce to the usual quantum
Langevin equations (\ref{qleqbm0})-(\ref{qleqbm}) in the limit 
$\Omega \rightarrow \infty$.
However, these equations must not be used as usual input-output
equations, as it is implicitely dictated by the SBMME of 
Eq.~(\ref{cl}) (see Ref.~\cite{JACOB}), but the appropriate 
commutation relations (\ref{secondo6}) and correlation functions 
(\ref{primo16})-(\ref{primo18}) must be used. Therefore the 
inadequacies found in Ref.~\cite{JACOB} are not due to the form of the 
Langevin equations but only to the inappropriate form of the 
noise correlation function dictated by the SBMME. 
It is also important to stress that 
our Langevin equation description of quantum Brownian 
motion is more general than that associated with a master equation 
approach, because it is valid {\em at all temperatures} and it does not
need any high temperature limit. 

Another criticism to the standard quantum Langevin equations 
presented in Ref.~\cite{JACOB} is that they do not preserve the commutation 
relations of system operators. This can also be traced back to
the inappropriate 
form of the commutation relations of the noise operator $\tilde{Q}(t)$.
Actually, Eqs.~(\ref{sblib}) and (\ref{secondo5}) are exact and,
because of the unitarity of the
time evolution, they mantain the initial commutation
rules for the system operators. This property is preserved also by 
the standard quantum Langevin equations (\ref{qleqbm0})-(\ref{qleqbm}), 
which are the 
$\Omega \rightarrow \infty $ limit of Eqs.~(\ref{sblib})-(\ref{secondo5}), 
provided 
that the correct noise commutation relation is used.
Let us see this fact in detail. If we do not restrict to the usual condition
$t > t_{0}$, under the $\Omega \rightarrow \infty $ limit, 
Eqs.~(\ref{sblib}) and (\ref{secondo5}) have to be written as (see Appendix) 
\begin{eqnarray}
\dot{q}(t) &=& p(t)/m \\
\dot{p}(t) &=& \frac{i}{\hbar} \left[H_{\cal S},p\right]+\tilde{Q}(t) 
 -2\eta \delta(t-t_{0}) q(t_{0}) - 
  \frac{\eta}{m} {\cal S}(t-t_{0}) p(t)  \; , 
\label{secondo7}
\end{eqnarray}
where ${\cal S}(t)$ is the sign function (defined so that
${\cal S}(0)=0$).

If we consider for example the commutator between $q(t)$ and $p(t)$, 
differentiate it with respect to $t$ and use Eqs.~(\ref{secondo7}), we obtain 
\begin{equation}
\frac{d}{d t} \big[ q(t), p(t) \big] = - \frac{\eta}{m}{\cal S}(t-t_{0})
\big[ q(t), p(t) \big] + \big[ q(t), \tilde{Q}(t) \big] \; .
\label{secondo8}
\end{equation}
In order to solve this equation we need the commutator between 
$q(t)$ and $\tilde{Q}(t)$. More in general we consider  
both quantities $
{\cal X}(t)=\big[ q(t), \tilde{Q}(t^{\prime}) \big]$ and
${\cal Y}(t)=\big[ p(t), \tilde{Q}(t^{\prime}) \big]$
as functions of $t$, with $t^{\prime}$ an independent parameter.
If for simplicity we restrict to the case of interest here
of a harmonically bound Brownian particle
$H_{\cal S}=m\omega_{\cal S}^2 q^2/2$,
it is possible to obtain the equation for $\dot{\cal X}(t)$ and 
$\dot{\cal Y}(t)$ 
using Eqs. (\ref{secondo7})
\begin{eqnarray}
\dot{\cal X}(t)&=&{\cal Y}(t)/m \label{secondo100} \\
\dot{\cal Y}(t)&=&-m\omega_{\cal S}^2 {\cal X}(t)
+2i \hbar \eta \delta^{\prime}(t-t^{\prime})- 
  \frac{\eta}{m} {\cal S}(t-t_{0}) {\cal Y}(t)  \; , 
\label{secondo10}
\end{eqnarray}
where we have used Eq.~(\ref{secondo6}) in the $\Omega \rightarrow 
\infty $ limit and the initial condition ${\cal X}(t_{0})=0$.
Using Eq.~(\ref{secondo7}), the 
solution of Eqs.~(\ref{secondo100})-(\ref{secondo10}) can be expressed in the 
following form:
\begin{eqnarray}
{\cal X}(t)&=& 2 \eta \frac{\partial}{\partial t^{\prime}} \left\{
\big[ q(t), q(t^{\prime}) \big] \, \Xi(t,t^{\prime},t_{0}) \right\} 
\label{secondo11} \\
{\cal Y}(t)&=& 2 \eta \frac{\partial}{\partial t^{\prime}} \left\{
\big[ p(t), q(t^{\prime}) \big] \, \Xi(t,t^{\prime},t_{0}) \right\} 
\; ,
\label{secondo12}
\end{eqnarray}
with 
\begin{equation}
\Xi(t,t^{\prime},t_{0})=\theta(t-t_{0})\theta(t-t^{\prime})
\theta(t^{\prime}-t_{0})-\theta(t_{0}-t)\theta(t^{\prime}-t)
\theta(t_{0}-t^{\prime}),
\label{secondo13}
\end{equation}
where $\theta(\tau)$ is the Heavyside step function (defined so that
$\theta(0)=1/2$).
Using Eqs.~(\ref{secondo7}) within Eq.~(\ref{secondo11}) 
and setting $t=t^{\prime}$, it is possible to observe that
\begin{equation}
{\cal X}(t^{\prime})=\big[ q(t^{\prime}), \tilde{Q}(t^{\prime}) \big]=
\frac{\eta}{m}{\cal S}(t^{\prime}-t_{0})
\big[ q(t^{\prime}), p(t^{\prime}) \big] .
\label{secondo14}
\end{equation}
Finally, using this results in Eq.~(\ref{secondo8}) with 
$t^{\prime}=t$, one gets the desired result, i.e.,
that the time derivative of the commutator 
between $q(t)$ and $p(t)$ is equal to zero.
Similar arguments can be
used to prove the preservation of all the other commutation relations
between Brownian particle operators.

\section{Homodyne spectrum}

Let us now consider again the dynamics of the optical mode
of the cavity with a movable mirror.
Using the results of the preceding section, the dynamics for $t> t_{0}$
of the system
can be described by the following set of coupled quantum Langevin equations
in the interaction picture with respect to $\hbar \omega_{0}b^{\dagger} b$ 
(see 
Eqs.~(\ref{hparte}),
(\ref{qle}), and (\ref{secondo7})),
\begin{eqnarray}
\dot{q}(t) &=& p(t)/m \label{terzo8uno}\\
\dot{p}(t) &=& -m \, \omega_{\cal S}^{2} q(t) + \tilde{Q}(t) -  
 \frac{\eta}{m} p(t)   +\frac{\hbar \omega_{c}}{L} b^{\dagger}(t) b(t)  \\
\dot{b}(t) &=& - \left(i \omega_{c} - i \omega_{0} +
\frac{\gamma_{c}}{2}\right) b(t) + i \frac{\omega_{c}}{L}q(t) b(t) +
E +
\sqrt{\gamma_{c}}b_{in}(t) \; , 
\label{terzo8}
\end{eqnarray}
where the commutation relations and correlation functions
of the input noise $b_{in}(t)$
are given respectively by Eqs.~(\ref{commu}), (\ref{correb1})
and (\ref{correb2}), while those of
$\tilde{Q}(t)$ are given by Eq.~(\ref{secondo6}) and  
Eqs.~(\ref{primo16})-(\ref{primo18}) considering the limit
$\Omega \rightarrow \infty$.

In standard interferometric applications, the driving field is
very intense. Under this condition the system is characterized by a
semiclassical steady state with the internal cavity mode in a coherent
state $|B_{st}\rangle $, and a new equilibrium position for the 
mirror, displaced by $q_{st} = \hbar \omega_{c}| B_{st} |^{2}/m 
\omega_{\cal S}^{2} L$ with respect to that with no driving field.
The steady state amplitude is given by the solution
of the nonlinear equation
\begin{equation}
	B_{st}=\frac{E}{\frac{\gamma_{c}}{2}+i \omega_{c} - i \omega_{0}
	- i \frac{\hbar \omega_{c}^{2}}{m \omega_{\cal S}^{2} 
	L^{2}}|B_{st}|^{2}} ,
	\label{nonlin}
\end{equation}
which is obtained by taking the expectation values of 
Eqs.~(\ref{terzo8uno})-(\ref{terzo8}), factorizing them and setting 
all the time derivatives to zero. Eq.~(\ref{nonlin}) shows a bistable 
behaviour which has been experimentally observed in \cite{dorsel}.

Under these semiclassical conditions, the dynamics is well described 
by linearizing the quantum Langevin equations 
(\ref{terzo8uno})-(\ref{terzo8}) around the steady state. If we now
rename with $q(t)$ and $b(t)$ the operators describing the quantum 
fluctuations around the classical steady state, one gets
\begin{eqnarray}
\dot{q}(t) &=& p(t)/m  \label{terzo12uno}\\
\dot{p}(t) &=& -m \omega_{\cal S}^{2} q(t) -  
  \frac{\eta}{m} p(t)   +\frac{\hbar \omega_{c}B_{st}}{L} \left( b(t) + 
  b^{\dagger}(t) \right) + \tilde{Q}(t) \\
\dot{b}(t) &=&  -\left(\frac{\gamma_{c}}{2} + i \Delta\right)b(t)
+i \frac{\omega_{c}B_{st}}{L}q(t)+ \sqrt{\gamma_{c}} b_{in}(t) \; ,
\label{terzo12bis}
\end{eqnarray} 
where we have chosen the phase of the cavity mode field so that
$B_{st}$ is real and 
\begin{equation}
	\Delta = \omega_{c} - \omega_{0} - 
	\frac{\hbar \omega_{c}^{2}}{m \omega_{\cal S}^{2} 
	L^{2}}|B_{st}|^{2}
	\label{detu}
\end{equation}
is the cavity mode detuning. We shall consider from now on $\Delta=0$, 
which corresponds to the most common experimental situation, and which can
always be achieved by appropriately adjusting the driving
field frequency $\omega_{0}$. In this case the dynamics becomes 
simpler, and
it is easy to see that 
only the phase quadrature $Y(t)=i\left(b^{\dagger}(t)-b(t)\right)$ is 
affected by the mirror position fluctuations $q(t)$,
while the amplitude field quadrature $X(t)=b(t)+b^{\dagger}(t)$ is not.
In particular, in the limit of a sufficiently large cavity mode 
bandwidth $\gamma_{c} \gg \eta/m , \omega_{\cal S}$ (which is
usually satisfied),  
the dynamics of the phase quadrature adiabatically follows that of the 
mirror position, that is
\begin{equation}
 Y(t) =\frac{4\omega_{c}
 B_{st}}{\gamma_{c}L} q(t)
+ \; {\rm noise \; terms.}
\end{equation}
Therefore a phase noise measurement, as for example the homodyne 
measurement of the field quadrature $Y(t)$, gives a direct information
on the mirror Brownian motion. In particular,
the interesting measurable quantity is the output power density spectrum of the
phase quadrature 
\begin{eqnarray}
S_{Y}(\omega) &=& \left\{ \int d\tau e^{i \omega \tau} \langle
 Y_{out}(t) Y_{out}(t+\tau)\rangle \right\}_{t} 
\label{terzo14} \\
&=&  \frac{1}{2\pi} \left\{ \int d \omega^{\prime} e^{-i (\omega+
\omega^{\prime}) t} 
\langle Y_{out}(\omega^{\prime})  
Y_{out}(\omega) \rangle 
\right\}_{t} \;,
\label{terzo15}
\end{eqnarray}
where $\{ \cdots \}_{t}$ denotes the time average over $t$, 
$Y_{out}(t) = i\left[b_{out}^{\dagger}(t)-b_{out}(t)\right]$ is the 
output phase quadrature,
$Y_{out}(\omega ) = \int dt e^{i\omega t} Y_{out}(t)$
is its
Fourier transform, and
\begin{equation}
b_{out}(t)+b_{in}(t)=\sqrt{\gamma_{c}}b(t)
\end{equation}
is the usual input-output relation \cite{milwal2}.
If we now take the Fourier transform of 
Eqs.~(\ref{terzo12uno})-(\ref{terzo12bis}), one gets the following 
expression for the Fourier transform of the correlation function
of the cavity mode annihilation operator
\begin{equation}
b(\omega) = \frac{1}{D(\omega)} \left[  i \frac{\omega_{c} 
B_{st}}{m L} \left( \frac{\hbar \omega_{c} B_{st}\sqrt{\gamma_{c}}}{L} 
\frac{b_{in}(\omega) + 
b_{in}^{\dagger}(\omega)}{i\omega-\gamma_{c}/2} -\tilde{Q}(\omega)\right) 
- \left(\omega_{\cal S}^{2} -\omega^{2} -i \eta 
\omega/m \right) \sqrt{\gamma_c}b_{in}(\omega) \right] \; ,
\label{terzo17b}
\end{equation}
with 
\begin{equation}
D(\omega) = \left( i \omega -\frac{\gamma_{c}}{2} \right) 
\left( \omega_{\cal S}^{2} -\omega^{2} -i \eta 
\omega/m  \right)  \; .
\label{primo17c}
\end{equation}

For the evaluation of the phase noise spectrum $S_{Y}(\omega )$,
one has to use Eq.~(\ref{terzo17b}) and then consider the 
spectrum of the various noise terms, which are easily derived 
from the Fourier transform of the correlation functions 
(\ref{correb1}), (\ref{correb2}), (\ref{primo16}), (\ref{primo17})
and (\ref{primo18}):
\begin{eqnarray}
\langle \tilde{Q}(\omega) \tilde{Q}(\omega^{\prime}) \rangle &=&
2 \pi \eta \hbar \omega \left[ 1 + \coth\left( \frac{\hbar \omega}{2 k T} 
\right) \right] \delta( \omega + \omega^{\prime}) \label{appec1} \\
\langle b_{in}^{\dagger}(\omega) b_{in}(\omega^{\prime}) \rangle &=&
0 \\
\langle b_{in}(\omega) b_{in}^{\dagger}(\omega^{\prime}) \rangle &=&
2\pi \delta( \omega + \omega^{\prime}) \;, \label{appec2}
\end{eqnarray}
where we have assumed again the infinite cutoff limit $\Omega 
\rightarrow \infty$ in the evaluation of the spectrum of the thermal 
noise operator $\tilde{Q}(t)$. One finally obtains 
\begin{equation}
S_{Y}(\omega) =
1+  4  \left( \frac{\hbar \omega_{c}^{2} \gamma_{c}|B_{st}|^{2}}{m L^{2}} \right)^{2} 
\frac{1}{ \left( (\gamma_{c}/2)^{2} + \omega^{2} \right) |D(\omega )|^{2} } 
+ 4 \left( \frac{ \omega_{c}^{2} 
\eta \gamma_{c} |B_{st}|^{2}}{m^{2} L^{2}} \right) \frac{1}{|D(\omega)|^{2}} 
\hbar \omega \coth \left(\frac{\hbar \omega}{2kT}\right) \; .
\label{terzo19}
\end{equation}
This is the phase noise spectrum associated with the homodyne 
measurement of the phase quadrature $Y(t)$, and the only assumptions 
made in its derivation are the linearization around the semiclassical 
steady state and the time coarse-grained description $\Omega 
\rightarrow \infty$. Its temperature dependence is instead exact and 
therefore Eq.~(\ref{terzo19}) is valid even at very low temperatures, 
differently from the spectra obtained with the approaches based on the 
master equation, as in Ref.~\cite{JACOB}, which cannot be applied in 
the low temperature limit.

Notice that $S_{Y}(\omega)$ is an even function of 
$\omega$, as it must be due to stationarity and the commutation rules
of output fields \cite{JACOB}. In fact, the non-sensical term of the 
spectrum found in Ref.~\cite{JACOB} in the case of the 
standard quantum Langevin description is due to the  
inappropriate form of the correlation function of the Langevin noise 
dictated by the SBMME, and it is absent when the correct spectrum of 
the quantum Brownian noise term of Eq.~(\ref{appec1}) is used.
This spectrum is plotted in Fig.~2
(see the figure caption for parameter values), where also the three 
contributions to the noise spectrum are explicitely shown. The full 
line refers to the total homodyne spectrum, while the
dashed-dotted line describes the shot noise, which is 
frequency-independent (actually $S_{Y}(\omega) $
has been defined so to be normalized just to the shot noise level).
The dashed line describes the second term in Eq.~(\ref{terzo19}), 
which is the one associated with the radiation pressure; finally the dotted 
line describes the last term which is just the thermal noise 
contribution.

The homodyne spectrum derived in Ref.~\cite{JACOB} with the adoption
of the Di\'osi master equation of Ref.~\cite{diosi1} coincides with
Eq.~(\ref{terzo19}) except for a different  
thermal noise term, which is obtained from that of Eq.~(\ref{terzo19})
with the replacement
\begin{equation}
\hbar \omega \coth\left(\frac{\hbar \omega}{2kT}\right) \rightarrow
2kT+\frac{\hbar^{2} 
\left(\omega^{2}+\eta^{2}/m^{2}\right)}{6kT} .
\label{compar}
\end{equation}
However, despite this formal difference, the two predictions become practically
indistinguishable if typical experimental parameters are considered.
In fact, the prediction of Ref.~\cite{JACOB} coincides with the high 
temperature expansion (at first order in $\hbar\omega/kT$) of 
Eq.~(\ref{terzo19}) except for the additional factor $\hbar \eta 
^{2}/6m^{2} kT$. However, in typical experiments, 
mechanical oscillators with a very good quality factor are always used, 
so that the 
term $\hbar^{2} \eta ^{2}/6m^{2} kT$ will be in practice always 
negligible with respect
to $\hbar^{2} \omega ^{2}/6 kT$ in Eq.~(\ref{compar}).
This means that an appreciable discrepancy between 
the two expressions of the thermal noise term manifests itself only
when $kT < \hbar \omega $, which means prohibitively small temperatures, or 
alternatively, very large frequencies (larger than 1 THz
at liquid He temperatures). Moroever at these high frequencies
the thermal noise contribution is completely blurred by the shot 
noise term and therefore we can conclude that with present tecnology
the phase noise spectrum of Eq.~(\ref{terzo19}) and that evaluated 
in Ref.~\cite{JACOB} cannot be experimentally distinguished.
Nonetheless, the result of Eq.~(\ref{terzo19}) is important because it
shows that the standard quantum Langevin 
equations (supplemented with the appropriate commutation relations 
(\ref{secondo6}) and correlation functions 
(\ref{primo16})-(\ref{primo18}) of the random Langevin force)
do give an adequate description of quantum Brownian motion, which 
is even more general than that associated with the master equation,
which is not valid at very low temperatures.

\section{Conclusions}

We have considered in this paper the dynamics of a cavity mode with a 
movable mirror, which is often used for the interferometric detection
of very weak forces. We have focused in particular on the description 
of the quantum Brownian motion of the mirror, which is responsible for 
the thermal noise term in the measured phase noise spectrum of the 
light reflected from the cavity. We have shown that
the standard quantum Langevin 
equations (\ref{qleqbm0})-(\ref{qleqbm}) provide an adequate and 
consistent description of quantum Brownian motion. We have derived the 
quantum Langevin equations directly from the independent oscillator 
model (providing the commonly used description for the oscillator 
reservoir, see \cite{caldleg,qnoise,caldleg2,ocon}), and we have seen 
that they provide a quite general description of quantum Brownian 
motion, valid at any temperatures. This is instead not true for master 
equation-based approaches, which cannot be applied in the 
low-temperature limit \cite{haake}. The inadequacies found in the 
quantum Langevin approach are to be traced back to the fact that the
quantum Langevin force appearing in it is different from the 
standard input noise terms of the input-output formalism \cite{io}, 
since it is characterized by a different commutation relation (see 
Eq.~(\ref{secondo6})) which does not coincide with a Dirac delta in 
any limit.

\appendix

\section{}

In order to justify the presence of the sign function ${\cal S}(t)$ on
Eq.~(\ref{secondo7}), let us consider
the step function
\begin{equation}
\theta(\tau)=\left\{ \begin{array}{rl}
   1 & \mbox{if $\tau>0$ } \\
   1/2 & \mbox{if $\tau=0$ } \\
   0 & \mbox{if $\tau<0$ } \; ,
 \end{array} \right.
\label{appea1}
\end{equation}
and the formal identity:
\begin{eqnarray}
{\cal I}(t,t_{0}) &=&
\int_{t_{0}}^{t} d t^{\prime} \delta(t-t^{\prime})p(t^{\prime}) 
\nonumber \\
&=& \int_{-\infty}^{\infty} d t^{\prime} \delta(t-t^{\prime}) p(t^{\prime})
\big( \theta(t^{\prime}-t_{0}) - \theta(t^{\prime}-t) \big) 
\label{appea2}
\end{eqnarray}
which holds for every $t$ and $t_{0}$. 
Now using the formal properties of $\delta(t)$ on Eq.~(\ref{appea2}),
one can verify that for $t>t_{0}$ 
and for $t<t_{0}$, ${\cal I}(t,t_{0})=p(t)/2$ and ${\cal I}(t,t_{0})=-p(t)/2$
respectively, while, of course, ${\cal I}(t_{0},t_{0})=0$. This 
can be written using the sign function ${\cal S}(\tau)= 
\theta(\tau)-\theta(-\tau)$ in the following way
\begin{equation}
{\cal I}(t,t_{0})={\cal S}(t-t_{0})p(t)/2 \; .
\label{appea3}
\end{equation}

\begin{figure}[htb]
\centerline{\epsfig{figure=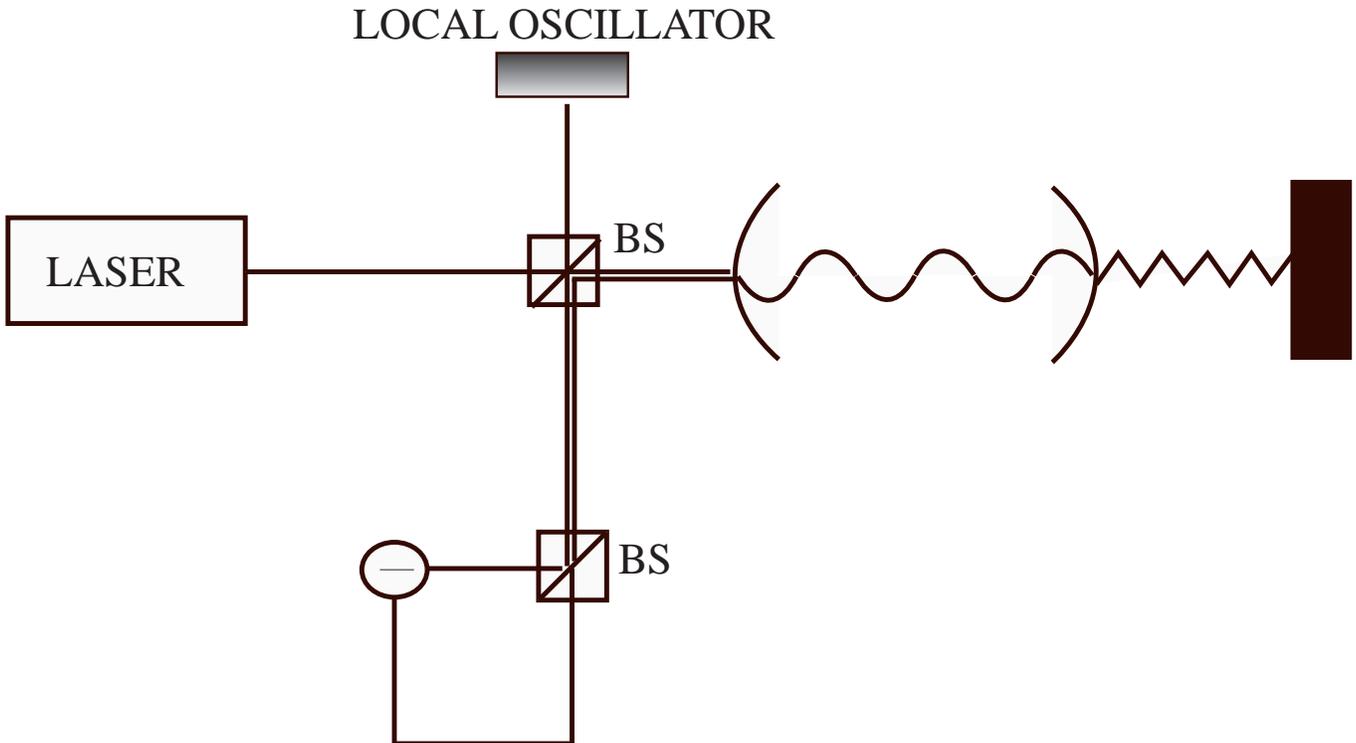,height=10cm}}
\caption{Schematical description of the system. The cavity mode
is driven by the laser which, thanks to the beam splitter BS,
provides also the local oscillator for the homodyne measurement of 
the light reflected by the cavity.}
\label{fig1}
\end{figure}

\begin{figure}[htb]

\centerline{\epsfig{figure=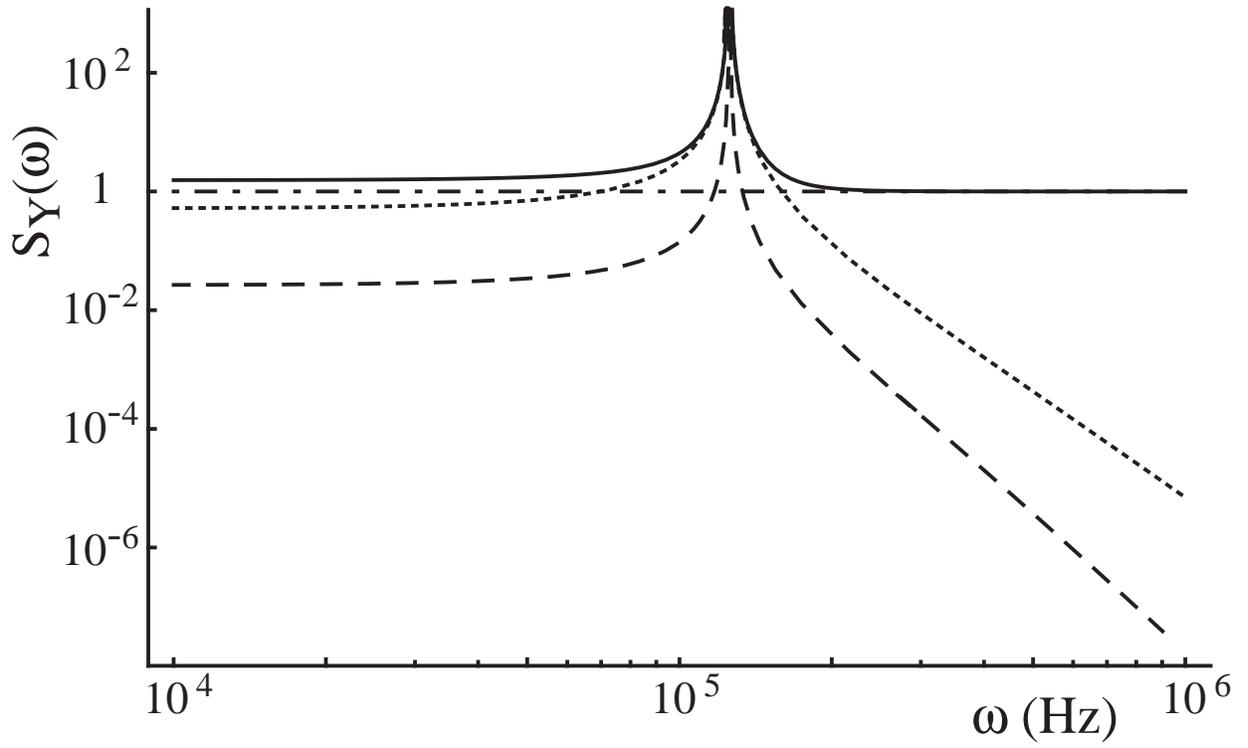,height=10cm}}
\caption{Phase noise spectrum of Eq.~(\protect\ref{terzo19})
(full line).
The dotted-dashed line refers to the shot noise spectrum (first term
of Eq.~(\protect\ref{terzo19})); the dashed line refers to the 
radiation pressure term (second term of Eq.~(\protect\ref{terzo19})),
and the dotted line to the thermal noise term (third term 
of Eq.~(\protect\ref{terzo19})). Parameter values are
$\omega_{\cal S}=1.3 \cdot 10^{5}$ Hz,  
$\eta/m = 3 \cdot 10^{-2}$ Hz, 
$\omega_{c}=1.8 \cdot 10^{15}$ Hz,
$\gamma_{c}=4.7 \cdot  10^{5}$ Hz, 
$m = 10^{-5}$ Kg, $L=10^{-2}$ m, 
$T=4.2 $ K, $P=10^{-5}$ W.}
\label{fig2}

\end{figure}

\end{document}